# A Trust-Based Cross-Layer Security Protocol for Mobile Ad hoc Networks

A.Rajaram
Anna University Coimbatore
Coimbatore, India
.

Dr.S.Palaniswami
Registrar, Anna University Coimbatore
Coimbatore, India

*Abstract*—In this paper, we develop a trust based security protocol based on a cross-layer approach which attains confidentiality and authentication of packets in both routing and link layers of MANETs. In the first phase of the protocol, we design a trust based packet forwarding scheme for detecting and isolating the malicious nodes using the routing layer information. It uses trust values to favor packet forwarding by maintaining a trust counter for each node. A node is punished or rewarded by decreasing or increasing the trust counter. If the trust counter value falls below a trust threshold, the corresponding intermediate node is marked as malicious. In the next phase of the protocol, we provide link-layer security using the CBC-X mode of authentication and encryption. By simulation results, we show that the proposed cross-layer security protocol achieves high packet delivery ratio while attaining low delay and overhead.

I. INTRODUCTION

*A. Mobile Ad-hoc Networks*

A mobile ad-hoc network (MANET) is a temporary infrastructure less multi-hop wireless network in which the nodes can move arbitrarily. Such networks extend the limited wireless transmission range of each node by multi-hop packet forwarding, thus, well suited for the scenarios in which pre deployed infrastructure support is not available. In an ad hoc network, there is no fixed infrastructure such as base stations or mobile switching centers. Mobile nodes that are within each other's radio range communicate directly via wireless links, while those that are far apart rely on other nodes to relay messages as routers. Node mobility in an ad hoc network causes frequent changes of the network topology. Mobile ad hoc networks are finding ever increasing applications in both military and civilian scenarios due to their self-organizing, self-configuring capabilities.

*B. Security Threats in MANETS*

An adhoc network can be attacked from any direction at any node which is different from the fixed hardwired networks with physical protection at firewall and gateways. Altogether it denotes that every node should be equipped to meet an attacker directly or indirectly.

Malicious attack can be initiated from both inside and outside of the network. Tracking a specific node is difficult in large adhoc networks and hence, it is more dangerous and much difficult to detect the attacks from an affected node. Altogether it denotes that every node should be prepared to work in a way that it should not trust on any node immediately.

Distributed architecture should be applied in order to achieve high availability. This is because if the central entity is used in the security solution, it causes serious attack on the entire network when the centralized entity gets affected.

The following are the types of active attacks and its relevant solutions:

**A. Black hole attack**

Let H be a malicious node. When H receives a Route Request, it sends back a Route Reply immediately, which constructs the data and can be transmitted by itself with the shortest path. So S receives Route Reply and it is replaced by H -> S. Then H receives all the data from S.

**B. Neighbor attack**

The neighbor attack and the black hole attack prevent the data from being delivered to the destination. But the neighbor attacker does not catch and capture the data packets from the source node. It leaves the settings as soon as sending the false messages.

**C. Wormhole attack**

Two malicious nodes share a private communication link between them. One node captures the traffic information of the network and sends them directly to other node. Warm hole can eavesdrop the traffic, maliciously drop the packets, and perform man-in- the-middle attacks against the network protocols. [6].

**D. DoS (Denial of Service) attack**

When the network bandwidth is hacked by a malicious node [5], then it results to the DoS attack. In order to utilize precious network resources like bandwidth, or to utilize node resources like memory or computation power, the attacker inserts packets into the network. The specific instances of the DoS attack are the routing table overflow attack and energy consumption attack.

**E. Information Disclosure attack**

The information disclosure attack aims at the privacy requirements of network. The confidential information's like routing location, node status or secret keys and password are leaked out by the malicious node to the unauthorized nodes.

### F. Rushing attack

The rushing attack aims against on-demand routing protocols which uses identical suppression at each node. In order to find routed to the destinations, the source nodes sends out the RREQ. Each intermediate node processes only the first non-duplicate packet and discards any duplicate packet which arrives at a later time. Rushing attackers can forward these packets quickly by skipping some of the routing processes. They are also able gain access to the forwarding group [7].

### G. Jellyfish attack

A malicious node receives and sends RREQ and RREP normally. But before forwarding it delays the data packets without any reason for some time [7]. Since the node has to intrude the forwarding group first, it is difficult to implement this type of attack. If the number of malicious node is few, then the influence to the network is also less.

### H. Byzantine attack

It is also called as impersonation attack because the malicious node might imitate another normal node. It also sends false routing information for creating an anomaly update in the routing table. In addition to this, an attacker may get unauthorized admission to resource and sensitive information.

### I. Blackmail attack

This attack is applicable against routing protocols which uses mechanisms for the recognition of malicious nodes and broadcast the messages which try to blacklist the offender [8]. By adding other legitimate nodes to their blacklists, an attacker might blackmail a legitimate node. Thus the nodes can be avoided in those routes.

## II. RELATED WORK

Farooq Anjum et al. [1] have proposed an initial approach to detect intrusions in ad hoc networks. Anand Patwardhan et al. [2] have proposed a secure routing protocol based on AODV over IPv6, further reinforced by a routing protocol-independent Intrusion Detection and Response system for ad-hoc networks. Chin-Yang Henry Tseng [3] has proposed a complete distributed intrusion detection system has consisted of four models for MANETs with formal reasoning.

Tarag Fahad and Robert Askwith [4] have concentrated on the detection phase and they have proposed a mechanism Packet Conservation Monitoring Algorithm (PCMA) is used to detect selfish nodes in MANETs. Panagiotis Papadimitratos and Zygmunt J. Haas[5] have proposed the secure message transmission (SMT) protocol and its alternative, the secure single-path (SSP) protocol SMT and SSP robustly detect transmission failures and continuously configure their operation to avoid and tolerate data loss, and to ensure the availability of communication. Ernesto Jiménez Caballero [6] has reviewed the possible attacks against the routing system, some of the IDSs proposed.

Yanchao Zhang et al. [7] have proposed a credit-based Secure Incentive Protocol (SIP) to stimulate cooperation in packet forwarding for infrastructure less MANETs. Liu et al. [8] have proposed the 2ACK scheme that has served as an add-on technique for routing schemes to detect routing misbehavior and to mitigate the adverse effect

Li Zhao and José G. Delgado-Frias [9] have proposed a scheme MARS and its enhancement E-MARS to detect misbehavior and mitigate adverse effects in ad hoc networks. Patwardhan et al. [10] have proposed an approach to secure a MANET using a threshold-based intrusion detection system and a secure routing protocol. Madhavi and Tai Hoon Kim [11] have proposed a MIDS (Mobile Intrusion Detection System) suitable for multi-hop ad-hoc wireless networks, which has detected nodes misbehavior, anomalies in packet forwarding, such as intermediate nodes dropping or delaying packets.

Syed Rehan Afzal et al. [12] have explored that the security problems and attacks in existing routing protocols and then they have presented the design and analysis of a secure on-demand routing protocol, called RSRP which confiscated the problems mentioned in the existing protocols. In addition, RSRP has used a very efficient broadcast authentication mechanism which does not require any clock synchronization and facilitates instant authentication

Bhalaji et al. [13] have proposed an approach based on the relationship between the nodes to make them to cooperate in an ad hoc environment. The trust values of each node in the network are calculated by the trust units. The relationship estimator has determined the relationship status of the nodes by using the calculated trust values. Their proposed enhanced protocol was compared with the standard DSR protocol and the results are analyzed using the network simulator-2.za

Kamal Deep Meka et al[14] have proposed a trust based framework to improve the security and robustness of adhoc network routing protocols. For constructing their trust framework they have selected the Ad hoc on demand Distance Vector (AODV) which is popular and used widely. Making minimum changes for implementing AODV and attaining increased level of security and reliability is their goal. Their schemes are based on incentives & penalties depending on the behavior of network nodes. Their schemes incur minimal additional overhead and preserve the lightweight nature of AODV.

Muhammad Mahmudul Islam et al. [15] have presented a possible framework of a link level security protocol (LLSP) to be deployed in a Suburban Ad-hoc Network (SAHN). They have analyzed various security aspects of LLSP to validate its effectiveness. To determine LLSP's practicability, they have estimated the timing requirement for each authentication process. Their initial work has indicated that LLSP is a suitable link-level security service for an ad-hoc network similar to a SAHN.

Shiqun Li et al. [16] have explored that the security issues of wireless sensor networks, and in particular propose an efficient link layer security scheme. To minimize computation and communication overheads of the scheme, they have designed a lightweight CBC-X mode Encryption/Decryption algorithm that attained encryption/decryption and authentication all in one. They have also devised a novel padding technique, enabling the scheme to achieve zero redundancy on sending encrypted/authenticated packets. As a result, security operations incur no extra byte in their scheme.

Stefan Schmidt et al. [17] have proposed security architecture for self-organizing mobile wireless sensor networks that prevented many attacks these networks are exposed to. In addition, it has limited the security impact of some attacks that cannot be prevented. They analyzed their security architecture and they have showed that it has provided the desired security aspects while still being a lightweight solution and thus being applicable for self-organizing mobile wireless sensor networks.

III. OBJECTIVES & OVERVIEW OF THE PROPOSED PROTOCOL

*A. Objectives*

In this paper, we propose to design a Trust-based Cross-layer Security protocol (TCLS) based on a cross-layer, approach which attains confidentiality and authentication of packets in routing layer and link layer of MANETs, having the following objectives:

- **lightweight** in order to considerably extend the network lifetime, that necessitates the application of ciphers that are computationally efficient like the symmetric-key algorithms and cryptographic hash functions
- **cooperative** for accomplishing high-level security with the aid of mutual collaboration/cooperation amidst nodes along with other protocols
- **attack-tolerant** to facilitate the network to resist attacks and device compromises besides assisting the network to heal itself by detecting, recognizing, and eliminating the sources of attacks;
- **flexible** enough to trade security for energy consumption;
- **compatible** with the security methodologies and services in existence
- **scalable** to the rapidly growing network size

*B. Overview of the Protocol*

We propose a Trust based packet forwarding scheme in MANETs without using any centralized infrastructure. It uses trust values to favor packet forwarding by maintaining a trust counter for each node. A node is punished or rewarded by decreasing or increasing the trust counter. Each intermediate node marks the packets by adding its hash value and forwards the packet towards the destination node. The destination node verifies the hash value and check the trust counter value. If the hash value is verified, the trust counter is incremented, otherwise it is decremented. If the trust counter value falls below a trust threshold, the corresponding the intermediate node is marked as malicious.

This scheme presents a solution to node selfishness without requiring any pre-deployed infrastructure. It is independent of any underlying routing protocol.

We focus on the CBC-X mode Encryption/Decryption algorithm to satisfy the necessity of minimum computational and communication overhead. This algorithm supports encryption/decryption and authentication of packets on a one-pass operation. The upper layers of the protocol stack are provided with security services obviously.

A CBC-X mode symmetric key mechanism is devised to employ our link layer security system. Encryption/Decryption and authentication operations are included into a single step which reduces the computational overhead to half, instead of calculating them individually. The padding technique states that this method has no cipher text expansion for the transmitted data payload. Thus the communication overhead is reduced significantly.

IV. EFFICIENT MAC LAYER SECURITY PROTOCOL

*A. Trust Based Forwarding Scheme*

In our proposed protocol, by dynamically calculating the nodes trust counter values, the source node can be able to select the more trusted routes rather than selecting the shorter routes. Our protocol marks and isolates the malicious nodes from participating in the network. So the potential damage caused by the malicious nodes are reduced. We make changes to the AODV routing protocol. An additional data structure called *Neighbors' Trust Counter Table (NTT)* is maintained by each network node.

Let $\{T_{c1}, T_{c2},......\}$ be the initial trust counters of the nodes $\{n_1, n_2,......\}$ along the route R1 from a source S to the destination D.

Since the node does not have any information about the reliability of its neighbors in the beginning, nodes can neither be fully trusted nor be fully distrusted. When a source S want to establish a route to the destination D, it send route request (RREQ) packets.

Each node keeps track of the number of packets it has forwarded through a route using a forward counter (FC). Each time, when node $n_k$ receives a packet from a node $n_i$, then $n_k$ increases the forward counter of node $n_i$.

$$FC_{n_i} = FC_{n_i} + 1, \quad i = 1,2..... \quad (1)$$

Then the NTT of node $n_k$ is modified with the values of $FC_{n_i}$.

Similarly each node determines its NTT and finally the packets reach the destination D.

When the destination D receives the accumulated RREQ message, it measures the number of packets received $Prec$. Then it constructs a MAC on $Prec$ with the key shared by the sender and the destination. The RREP contains the source and destination ids, The MAC of $Prec$, the accumulated route from the RREQ, which are digitally signed by the destination. The RREP is sent towards the source on the reverse route R1.

Each intermediate node along the reverse route from D to S checks the RREP packet to compute success ratio as,

$$SR_i = FC_{n_i} / Prec \qquad (2)$$

Where $Prec$ is the number of packets received at D in time interval $t_1$. The $FC_{n_i}$ values of $n_i$ can be got from the corresponding NTT of the node. The success ratio value $SR_i$ is then added with the RREP packet.

The intermediate node then verifies the digital signature of the destination node stored in the RREP packet, is valid. If the verification fails, then the RREP packet is dropped. Otherwise, it is signed by the intermediate node and forwarded to the next node in the reverse route.

When the source S receives the RREP packet, if first verifies that the first id of the route stored by the RREP is its neighbor. If it is true, then it verifies all the digital signatures of the intermediate nodes, in the RREP packet. If all these verifications are successful, then the trust counter values of the nodes are incremented as

$$Tc_i = Tc_i + \delta 1 \qquad (3)$$

If the verification is failed, then

$$Tc_i = Tc_i - \delta 1 \qquad (4)$$

Where $\delta 1$ is the step value, which can be assigned a small fractional value during the simulation experiments.

After this verification stage, the source S check the success ratio values $SR_i$ of the nodes $n_i$.

For any node $n_k$, if $SR_k < SR_{min}$, where $SR_{min}$ is the minimum threshold value, its trust counter value is further decremented as

$$Tc_i = Tc_i - \delta 2 \qquad (5)$$

For all the other nodes with $SR_k > SR_{min}$, the trust counter values are further incremented as

$$Tc_i = Tc_i + \delta 2 \qquad (6)$$

Where $\delta 2$ is another step value with $\delta 2 < \delta 1$.

For a node $n_k$, if $T_{c_k} < T_{c_{thr}}$, where $T_{c_{thr}}$ is the trust threshold value, then that node is considered and marked as malicious.

If the source does not get the RREP packet for a time period of $t$ seconds, it will be considered as a route breakage or failure. Then the route discovery process is initiated by the source again.

The same procedure is repeated for the other routes R2, R3 etc and either a route without a malicious node or with least number of malicious nodes, is selected as the reliable route.

In this protocol, authentication is performed for route reply operation. Also, only nodes which are stored in the current route, need to perform these cryptographic computation. So the proposed protocol is efficient and more secure.

*B. CBC-X Mode*

Our proposed link layer security scheme adapts the packet format of [16]. But the encryption and decryption mechanisms are different.

It works between the link layer and the radio layer. Our proposed method encrypts the data and computes the MAC, when the application data payload is passed from the link layer to the radio layer. With the help of the radio channel, the encrypted message is sent out bit-by-bit. Confidentiality and authentication are the of security services which are present in our proposed packet format.

The packet format of the proposed scheme is illustrated in Figure.1. The fields of the packet are the destination address field, the active message type field, the length field and the data field. We keep the one byte group field in the proposed scheme to make it general and applicable. We also use a 4 byte MAC field since it can provide enough security of integrity and authenticity for the mobile adhoc networks. Any error alteration during message transmission can be detected by re-computing the MAC and the error message would be discarded to improve efficiency. It uses an 8 byte initial vector (IV) and a block cipher mechanism to encrypt the data field of the packet. The fixed portions of both IVs are the destination address field, the link type field and the length field. These fields take 4 bytes totally.

| Dest | A | L | G | Ran | Data | MAC |
|------|---|---|---|-----|------|-----|
| 2    | 1 | 1 | 1 | 3   | (0-29) | 4 |

Figure 1. Packet Format

In our scheme, the generic communication interfaces are given to the upper layer and uses the lower radio packet interfaces. The nodes in the communication are not conscious of the operations on encryption/authentication because the security services are given clearly. To make the scheme easier, the encryption and authentication for every packet is carried out by our default mode in a single pass. In order to finish the message authentication and encryption concurrently before sending message, we built an authentication and encryption scheme called as CBC-X mode.

*1) CBC-X Mode Operations:*

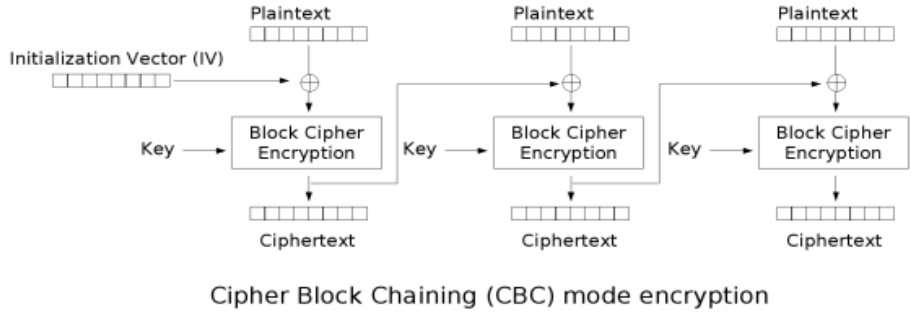

Figure 2. Encryption

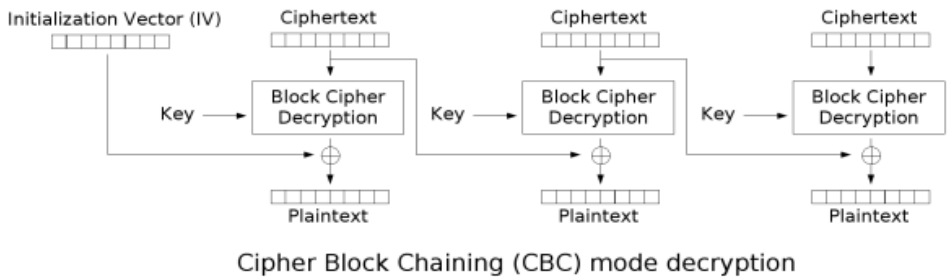

Figure 3. Decryption

The basic steps involved in the encryption and decryption operations are illustrated in figure 2 and figure 3, respectively.

If the first block has index 1, the formula for CBC encryption is $C_i = E_K(P_i \oplus C_{i-1}), \quad C_0 = IV$

while the formula for CBC decryption is $P_i = D_K(C_i) \oplus C_{i-1}), \quad C_0 = IV$

The working of the present CBC mode is described below: One cipher text block will be returned for each plaintext block, if a part of the plaintext is encrypted. In encryption of the last block of the plaintext, one or two cipher text blocks can be returned. On the other hand, decryption works in the reverse order. Apart from the decryption of the last block, a one plaintext block will be returned for each cipher text block. After the decryption of the last plaintext block, its padding is calculated and cut off, returning a valid plaintext.

*2) CBC Padding Schemes:* Plaintext is divided into blocks of 8 bytes (64 bits). The final plaintext block must be padded: the final $a$ plaintext bytes $0 \leq a \leq 7$ are followed by $8-a$ padding bytes, valued $8-a$.
For example:
*messagebyte1|| messagebyte2 ||'06'||'06'||'06'||'06'||'06'||'06'  ESP*

$X$ *padding bytes* $1 \leq X \leq 255$

*'01'||'02'||'03'||…..||' X'*

## V. PERFORMANCE EVALUATION

### A. Simulation Model and Parameters

We use NS2 to simulate our proposed algorithm. In our simulation, the channel capacity of mobile hosts is set to the same value: 2 Mbps. We use the distributed coordination function (DCF) of IEEE 802.11 for wireless LANs as the MAC layer protocol. It has the functionality to notify the network layer about link breakage.

In our simulation, 100 mobile nodes move in a 1000 meter x 1000 meter square region for 50 seconds simulation time. We assume each node moves independently with the same average speed. All nodes have the same transmission range of 250 meters. In our simulation, the speed is varied from 10 m/s to 50m/s. The simulated traffic is Constant Bit Rate (CBR).

Our simulation settings and parameters are summarized in table I

TABLE I. SIMULATION PARAMETERS

| No. of Nodes | 100 |
|---|---|
| Area Size | 1000 X 1000 |
| Mac | 802.11 |
| Radio Range | 250m |
| Simulation Time | 50 sec |
| Traffic Source | CBR |
| Packet Size | 512 |
| Mobility Model | Random Way Point |
| Speed | 10,20,30,40,50m/s |
| Pause time | 5 |

## B. Performance Metrics

We evaluate mainly the performance according to the following metrics.

**Control overhead:** The control overhead is defined as the total number of routing control packets normalized by the total number of received data packets.

**Average end-to-end delay:** The end-to-end-delay is averaged over all surviving data packets from the sources to the destinations.

**Average Packet Delivery Ratio:** It is the ratio of the number .of packets received successfully and the total number of packets transmitted.

The simulation results are presented in the next section. We compare our TCLS protocol with the LLSP [15] protocol in presence of malicious node environment.

## C. Results

### A. Based On Attackers

In our First experiment, we vary the no. of misbehaving nodes as 5,10,15,20 and 25.

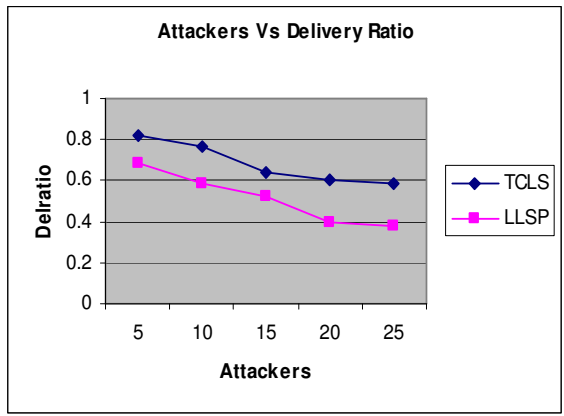

Figure 4. Attackers Vs Delivery Ratio

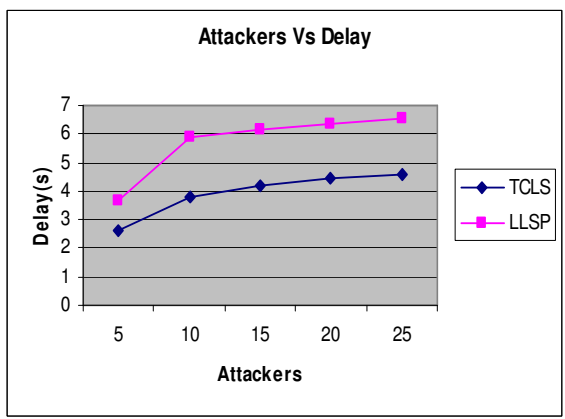

Figure 5. Attackers Vs Delay

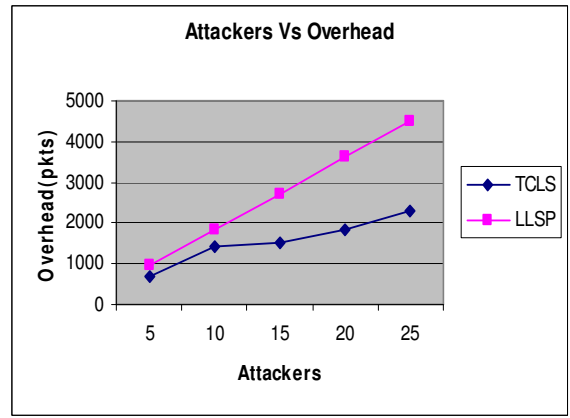

Figure 6. Attackers Vs Overhead

Figure 4 show the results of average packet delivery ratio for the misbehaving nodes 5, 10….25 scenario. Clearly our TCLS scheme achieves more delivery ratio than the LLSP scheme since it has both reliability and security features.

Figure 5 shows the results of average end-to-end delay for the misbehaving nodes 5, 10….25. From the results, we can see that TCLS scheme has slightly lower delay than the LLSP scheme because of authentication routines.

Figure 6 shows the results of routing overhead for the misbehaving nodes 5, 10….25. From the results, we can see that TCLS scheme has less routing overhead than the LLSP scheme since it does not involve route re-discovery routines.

### B. Based On Speed

In our Second experiment, we vary the speed as 10,20,30,40 and 50, with 5 attackers.

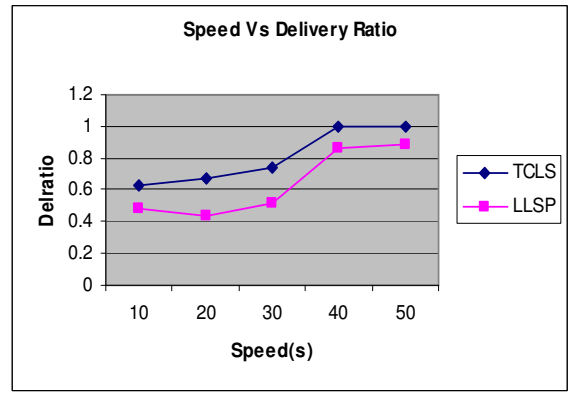

Figure 7. Speed Vs Delivery Ratio

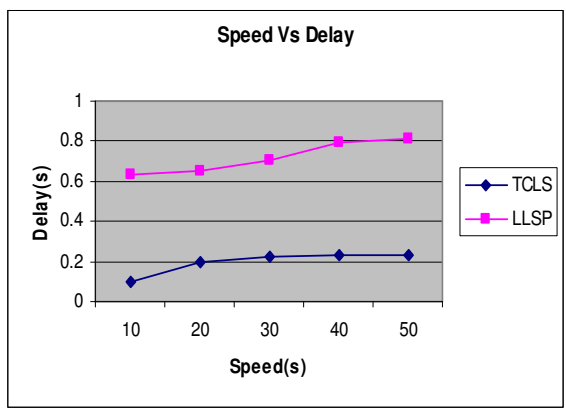

Figure 8. Speed Vs Delay

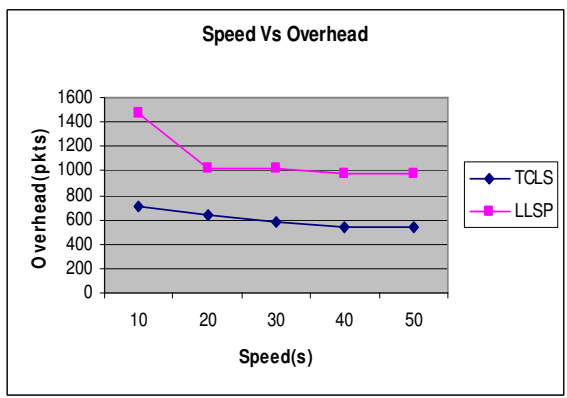

Figure 9. Speed Vs Overhead

Figure 7 show the results of average packet delivery ratio for the speed 10, 20…50 for the 100 nodes scenario. Clearly our TCLS scheme achieves more delivery ratio than the LLSP scheme since it has both reliability and security features.

Figure 8 shows the results of average end-to-end delay for the speed10, 20….50. From the results, we can see that TCLS scheme has slightly lower delay than the LLSP scheme because of authentication routines

Figure 9 shows the results of routing overhead for the speed 10, 20….50. From the results, we can see that TCLS scheme has less routing overhead than the LLSP scheme.

VI. CONCLUSION

In this paper, we have developed a trust based security protocol which attains confidentiality and authentication of packets in both routing and link layers of MANETs. In the first phase of the protocol, we have designed a trust based packet forwarding scheme for detecting and isolating the malicious nodes using the routing layer information. It uses trust values to favor packet forwarding by maintaining a trust counter for each node. A node is punished or rewarded by decreasing or increasing the trust counter. If the trust counter value falls below a trust threshold, the corresponding intermediate node is marked as malicious. In this protocol, authentication is performed for route reply operation. Also, only nodes which are stored in the current route need to perform this cryptographic computation. So the proposed protocol is efficient and more secure. This scheme presents a solution to node selfishness without requiring any pre-deployed infrastructure. It is independent of any underlying routing protocol. In the next phase of the protocol, we provide link-layer security using the CBC-X mode of authentication and encryption. By simulation results, we have shown that the proposed cross-layer security protocol achieves high packet delivery ratio while attaining low delay and overhead. As a future work, we will try to reduce the energy consumption, control overhead and delay of our proposed protocol by applying some optimization techniques.

REFERENCES

[1] Farooq Anjum, Dhanant Subhadrabandhu and Saswati Sarkar "Signature based Intrusion Detection for Wireless Ad-Hoc Networks: A Comparative study of various routing protocols" in proceedings of IEEE 58th Conference on Vehicular Technology, 2003.

[2] Anand Patwardhan, Jim Parker, Anupam Joshi, Michaela Iorga and Tom Karygiannis "Secure Routing and Intrusion Detection in Ad Hoc Networks" Third IEEE International Conference on Pervasive Computing and Communications, March 2005.

[3] Chin-Yang Henry Tseng, "Distributed Intrusion Detection Models for Mobile Ad Hoc Networks" University of California at Davis Davis, CA, USA, 2006.

[4] Tarag Fahad and Robert Askwith "A Node Misbehaviour Detection Mechanism for Mobile Ad-hoc Networks", in proceedings of the 7th Annual PostGraduate Symposium on The Convergence of Telecommunications, Networking and Broadcasting, June 2006.

[5] Panagiotis Papadimitratos, and Zygmunt J. Haas, "Secure Data Communication in Mobile Ad Hoc Networks", IEEE Journal On Selected Areas In Communications, Vol. 24, No. 2, February 2006.

[6] Ernesto Jiménez Caballero, "Vulnerabilities of Intrusion Detection Systems in Mobile Ad-hoc Networks - The routing problem", 2006.

[7] Yanchao Zhang, Wenjing Lou, Wei Liu, and Yuguang Fang, "A secure incentive protocol for mobile ad hoc networks", *Wireless Networks (WINET)*, vol 13, No. 5, October 2007.

[8] Liu, Kejun Deng, Jing Varshney, Pramod K. Balakrishnan and Kashyap "An Acknowledgment-based Approach for the Detection of Routing Misbehavior in MANETs", IEEE Transactions on Mobile Computing, May 2007.

[9] Li Zhao and José G. Delgado-Frias "MARS: Misbehavior Detection in Ad Hoc Networks", in proceedings of IEEE Conference on Global Telecommunications Conference,November 2007.

[10] A.Patwardhan, J.Parker, M.Iorga, A. Joshi, T.Karygiannis and Y.Yesha "Threshold-based Intrusion Detection in Adhoc Networks and Secure AODV" Elsevier Science Publishers B. V. , Ad Hoc Networks Journal (ADHOCNET), June 2008.

[11] S.Madhavi and Dr. Tai Hoon Kim "An Intrusion Detection System In Mobile Adhoc networks" International Journal of Security and Its Applications Vol. 2, No.3, July, 2008.

[12] Afzal, Biswas, Jong-bin Koh,Raza, Gunhee Lee and Dong-kyoo Kim, "RSRP: A Robust Secure Routing Protocol for Mobile Ad Hoc Networks", in proceedings of IEEE Conference on Wireless Communications and Networking, pp.2313-2318,April 2008.

[13] Bhalaji, Sivaramkrishnan, Sinchan Banerjee, Sundar, and Shanmugam, "Trust Enhanced Dynamic Source Routing Protocol for Adhoc Networks", in proceedings of World Academy Of Science, Engineering And Technology, Vol. 36, pp.1373-1378, December 2008

[14] Meka, Virendra, and Upadhyaya, "Trust based routing decisions in mobile ad-hoc networks" In Proceedings of the Workshop on Secure Knowledge Management, 2006.

[15] Muhammad Mahmudul Islam, Ronald Pose and Carlo Kopp, "A Link Layer Security Protocol for Suburban Ad-Hoc Networks", in proceedings of Australian Telecommunication Networks and Applications Conference, December 2004.

[16] Shiqun Li, Tieyan Li, Xinkai Wang, Jianying Zhou and Kefei Chen, "Efficient Link Layer Security Scheme for Wireless Sensor Networks", Journal of Information And Computational Science, Vol.4, No.2,pp. 553-567, June 2007.

[17] S. Schmidt, H. Krahn, S. Fischer, and D. Wätjen, "A Security Architecture for Mobile Wireless Sensor Networks", In proceedings of First European Workshop on Security in Ad-Hoc and Sensor Networks (ESAS 2004), August 2004.

AUTHORS PROFILE

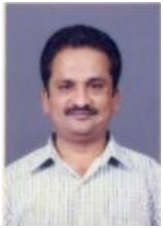

**S. Palaniswami** received the **B.E.** degree in electrical and electronics engineering from the Govt., college of Technology, Coimbatore, University of Madras, Madras, India, in 1981, the **M.E.** degree in electronics and communication engineering (Applied Electronics) from the Govt., college of Technology, Bharathiar University, Coimbatore, India, in 1986 and the **Ph.D.** degree in electrical engineering from the **PSG** Technology, Bharathiar University, Coimbatore, India, in 2003. He is currently the Registrar of Anna University Coimbatore, Coimbatore, India, Since May 2007. His research interests include Control systems, Communication and Networks, Fuzzy logic and Networks, **AI**, Sensor Networks. . He has about 25 years of teaching experience, since 1982. He has served as lecturer, Associate Professor, Professor, Registrar and the life Member of **ISTE**, India.

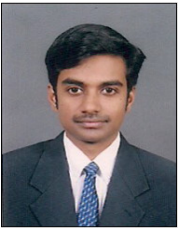

**A. Rajaram** received the **B.E.** degree in electronics and communication engineering from the Govt., college of Technology, Coimbatore, Anna University, Chennai, India, in 2006, the **M.E.** degree in electronics and communication engineering (Applied Electronics) from the Govt., college of Technology, Anna University, Chennai, India, in 2008 and he is currently pursuing the full time **Ph.D.** degree in electronics and communication engineering from the Anna University Coimbatore, Coimbatore, India. His research interests include communication and networks mobile adhoc networks, wireless communication networks **(WiFi, WiMax HighSlot GSM),** novel **VLSI NOC** Design approaches to address issues such as low-power, cross-talk, hardware acceleration, Design issues includes **OFDM MIMO** and noise Suppression in **MAI** Systems, **ASIC** design, Control systems, Fuzzy logic and Networks, **AI**, Sensor Networks.